\documentclass[a4paper, amsfonts, amssymb, amsmath, reprint, showkeys, nofootinbib, twoside]{revtex4-1}
\usepackage[english]{babel}
\usepackage[utf8]{inputenc}
\usepackage{subcaption}
\usepackage[colorinlistoftodos, color=green!40, prependcaption]{todonotes}
\usepackage{amsthm}
\usepackage{mathtools}
\usepackage{physics}
\usepackage{xcolor}
\usepackage{graphicx}
\usepackage[left=23mm,right=13mm,top=35mm,columnsep=15pt]{geometry} 
\usepackage{adjustbox}
\usepackage{placeins}
\usepackage[T1]{fontenc}
\usepackage{lipsum}
\usepackage{csquotes}
\newcommand{\energy}[1]{$\sqrt{s_{\text{NN}}}$ = #1 TeV}
\newcommand{\two}[2]{$#1_\text{#2}$}
\newcommand{\mtwo}[2]{#1_\text{#2}}
\newcommand{\fq}[1]{$F_\text{#1} \text{(M)}$}

\newcommand{\sqm}[0]{$M^\text{2}$}
\newcommand{\ptrange}[2]{$#1 \leq p_\text{T} \leq #2$}

\usepackage[pdftex, pdftitle={Article}, pdfauthor={Author}]{hyperref} % For hyperlinks in the PDF
\begin{document}
\title{Intermittency and fractal behaviour of charged particles in EPOS4 and PYTHIA8 generated events at LHC energies}
\author{Fakhar Ul Haider}
\email{fakharulhaider@gmail.com}
\author{Ramni Gupta}
\email{ramni.gupta@cern.ch}
\author{Salman Khurshid Malik}
\email[]{salmankm@keemail.me}
\author{Balwan Singh}
%\email{balwan.singh@cern.ch}
\author{Zarina Banoo}
%\email{zarina.banoo@cern.ch}
\author{Pryianka Choudhary}
%\email{pryiankachoudhary26@gmail.com}
\affiliation{Department of Physics, University of Jammu, Jammu, J\&K, India} 
\date{\today} % Leave empty to omit a date

\begin{abstract}
Large number density fluctuations of the charged particles produced in heavy-ion collisions are a promising signature for exploring the QCD phase transition and critical point in the nuclear matter phase diagram. Intermittency methodology is used to probe the fractal and scale invariant nature of these fluctuations. Intermittency is the phenomenon of power-law growth of the normalized factorial moments ($F_{\rm{q}}$) of the number density distributions over decreasing bin size. The charged particles  generated in the midrapidity region using PYTHIA8 and EPOS4 (UrQMD ON/OFF) for Pb--Pb collisions at $\sqrt{s_{\text{NN}}}$ = 5.02 TeV are studied. Scaling behaviour of $F_{\rm{q}}$ are studied as a function of phase space partitioning and second order moments to quantify the particle production nature within the default constraints of the two models. The scaling exponent related to the phase transition and parameters connected to fractal nature obtained for both these models show the absence of fluctuations of critical nature and multifractal behaviour.
\end{abstract}

\maketitle 
\section{Introduction} \label{sec:introduction}
The Quark Gluon Plasma (QGP)~\cite{Gribov:1972ri}, a deconfined state of quarks and gluons is one of the most intriguing frontiers in high-energy physics. Believed to have existed a few microseconds after the Big Bang~\cite{Weinberg:1977ji}, the QGP created in the heavy-ion collisions offers a unique opportunity to study the fundamental dynamics of Quantum Chromodynamics (QCD)~\cite{Barber:1979yr} under extreme conditions. QCD predicts that, at sufficiently high temperature and baryon density, strongly interacting matter undergoes a phase transition from hadronic matter to a deconfined  state of matter (QGP)~\cite{Shuryak:1978ij, Niida:2021wut, PHENIX:2003pfh}. During the last two decades confirmed experimental signatures of QGP formation in the heavy-ion collisions have been observed. Now mapping the QCD phase diagram and locating the critical point is one of the central goals in the field. Experiments at the Large Hadron Collider (LHC)~\cite{Lyndon} at CERN, Geneva and Relativistic Heavy Ion Collider (RHIC) at BNL in US,  recreate these extreme conditions through ultra-relativistic heavy-ion collisions, producing a small short-lived but extremely hot and dense medium that rapidly expands, cools and hadronizes. A large number of particles are produced during these collision experiments that provide valuable insight into the properties of QGP at extreme conditions. Fluctuations in the number of these produced particles called multiplicity fluctuations~\cite{Baym:1999up,Koch:2001zn}, is one of the many observables which are used in the field to perform studies to get answers to many unknown aspects of the nature of matter at its fundamentals. 
\par
Multiplicity fluctuations are a sensitive probe to the underlying particle production mechanism and to the dynamics of the system involved~\cite{Kliemant:2008rh,Stephanov:1999zu}.  By studying scaling behaviour of the normalized factorial moments (NFMs) of number-density distributions~\cite{Hwa:2011bu} of produced particles, insights into the underlying dynamics of the QCD phase transition can be obtained. The power--law scaling behaviour of the NFM with decreasing bin size, termed as intermittency, is considered as a promising tool for investigating the dynamics of the systems. Studies of heavy-ion collisions suggest the presence of multifractal behaviour~\cite{Sarkisyan:2015gca, Borghini:2003ur} in the system formed during these collisions. The anomalous fractal dimension is found not to remain constant but instead to depend on the order of the moment through a universal scaling relation applicable to systems described by formalism for second-order phase transition under Ginzburg--Landau (GL) theory~\cite{Hwa:1992uq}. The unprecedented collision energies and high particle multiplicities available at the experimental facilities performing heavy-ion collisions at ultra-relativistic energies provide a rich dataset for exploring these phenomena in greater detail. Such studies offer the potential to reveal new physics and further improve our understanding of the underlying mechanisms of QCD, while also providing stringent tests and refinements of existing theoretical models. 
\\
This article presents a study of intermittency and fractal behaviour using events generated with PYTHIA8+Angantyr~\cite{Sjostrand:2014zea} and EPOS4~\cite{Werner:2024fwk} in two different modes. Section II gives a brief overview of these Monte Carlo event generators and the event samples used in the analysis. The analysis methodology is described in Section III, followed by observations and results being given in section IV. Section V summarizes the main findings of this work.

\section{Monte Carlo event generators and events samples} \label{sec:eventgenerators}
\subsection{Monte Carlo event generators}
PYTHIA8 is a general-purpose Monte Carlo event generator designed primarily for simulating high-energy proton--proton $(pp)$ collisions. It provides a complete description of the event evolution, starting from the hard partonic scatterings to final-state hadron production. A key feature of PYTHIA8 ~\cite{Sjostrand:2014zea} is its detailed treatment of QCD parton scatterings through an extensive set of internally implemented ${(2 \rightarrow 1)}$ and ${(2 \rightarrow 2)}$ hard processes, complemented by multi-parton interaction (MPI) modeling ~\cite{Sjostrand:1987su}, where several independent QCD scatterings can occur within a single event. These scatterings are regulated using an energy-dependent transverse momentum cutoff to avoid divergences at low ${p_{T}}$. The probability for multiple interactions is determined based on parton distribution functions and an overlap function that depends on the impact parameter of the collision. Colour reconnection schemes~\cite{Christiansen_2014} limit the number of independent colour strings, thereby preventing an overproduction of soft particles at large rapidities and shapes the rapidity profile of final state particles.

\begin{figure}[htp]
\centering
\includegraphics[scale=0.45]{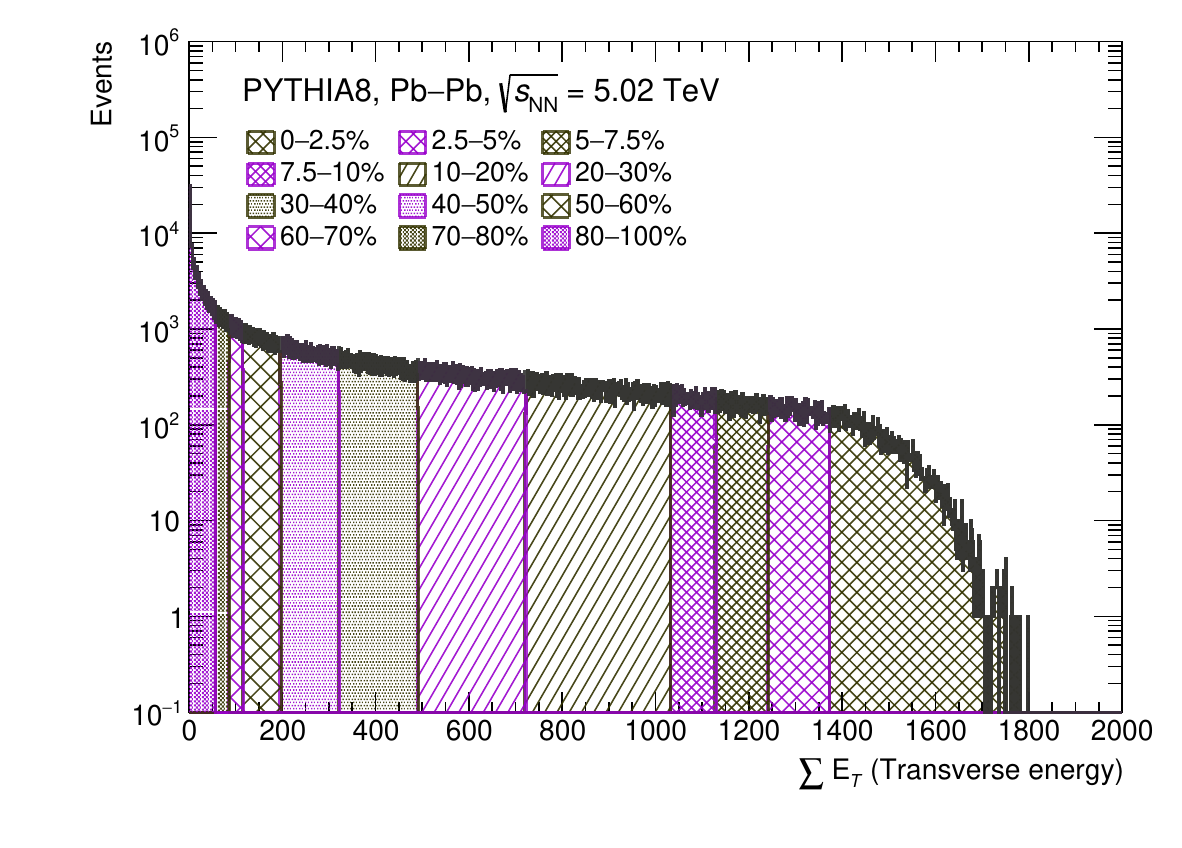}
\caption{Distribution of the summed transverse energy, $\sum \mtwo{E}{T}$ used to define centrality classes for PYTHIA8 for Pb--Pb collisions at \energy{5.02}. Centrality intervals are shown by percentile boundaries.}
\label{figure3}
\end{figure}
\begin{figure}[htp]
\includegraphics[scale=0.45]{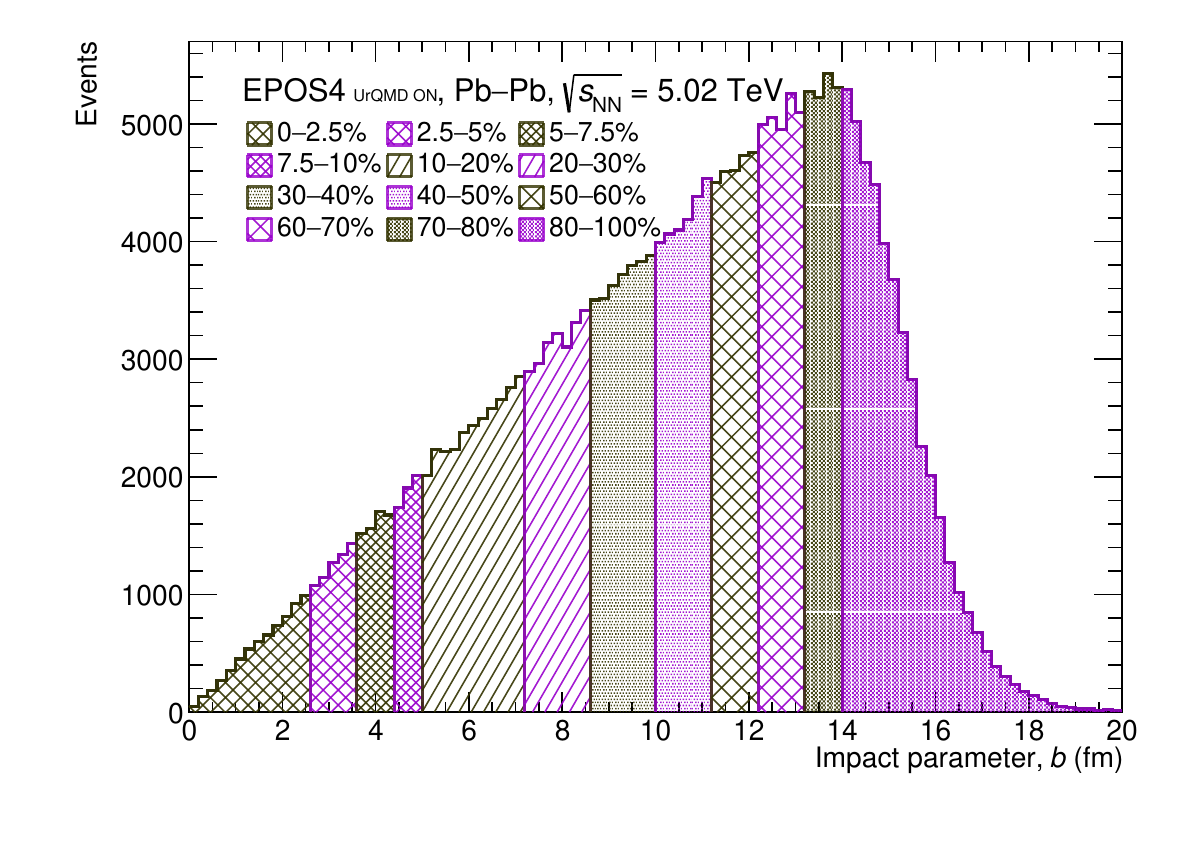}
\caption{Impact-parameter distribution for Pb--Pb collisions at \energy{5.02} using EPOS4. Vertical boundaries in the plot indicate centrality percentiles.}
\label{figure2}
\end{figure}
To simulate more complex systems such as proton--nucleus (pA) and nucleus--nucleus (AA) collisions, the Angantyr~\cite{Bierlich_2018} model extends PYTHIA8  by embedding it within a Glauber-inspired~\cite{Glauber:1955qq} framework. In this approach, the number of binary nucleon--nucleon (NN) sub-collisions is determined from the geometry of the collision using Monte Carlo Glauber calculations~\cite{Miller_2007}. In AA collisions, additional complexities such as energy--momentum conservation, impact parameter reweighing \cite{Mallick:2021wop}, and isospin adjustments are implemented to better reflect nuclear dynamics. Within this combined PYTHIA--Angantyr framework, multiplicity fluctuations originate first from geometric fluctuations that emerge from variations in the number and configuration of NN sub-collisions from event-to-event and second from internal fluctuations that emerge from PYTHIA’s MPI model, which governs the number of partonic interactions within each NN sub-event. In pA and AA collisions ~\cite{Ivanyi:1999bv,Pi:1992ug}, these fluctuations are amplified due to the presence of secondary interactions and the varying energy available in the nucleon remnants. The interplay between these sources result in realistic multiplicity distributions and provides a reliable baseline for interpreting initial-state fluctuations, non-flow correlations and collective behaviour in heavy-ion collisions.
\par
The second MC model that is studied is EPOS4~\cite{Werner:2023fne} which is an event generator framework developed for simulating high-energy collisions like proton-proton (pp) and nucleus-nucleus (AA) interactions. It employs a rigorous parallel scattering  framework that simultaneously models multiple parton–parton interactions within a single collision event, representing a significant advancement over traditional QCD-based event generators. Within this model, particle production is organized through a core–corona separation, in which the system formed at early times is decomposed into regions of high and low density. The high-density component gives rise to a collectively expanding medium described by hydrodynamics, while particles originating from the low-density component escape largely unaffected and hadronize without collective behaviour. This scheme provides a natural interpolation between thermalized and non-thermalized particle production across different collision systems. In addition, EPOS4 allows the inclusion of late-stage hadronic interactions by coupling to the UrQMD transport model. Simulations with UrQMD enabled EPOS4 incorporate hadronic rescattering and resonance decays, whereas calculations without UrQMD effectively exclude the hadronic phase, offering a transparent way to quantify the influence of hadronic interactions on final-state observables~\cite{Bleicher:1999xi}.

Key strength of EPOS4 lies in its ability to compute dedicated parton distribution functions and inclusive cross sections, facilitating accurate descriptions of hard probes like jets in conjunction with the underlying soft background and global particle production. It is capable of modelling charged-particle production for various collision systems and centrality classes, thereby providing a detailed understanding of the mechanisms governing particle generation, hadronization and subsequent final-state dynamics~\cite{Werner:2023zvo}. 
\begin{figure}
\centering
\includegraphics[scale=0.45]{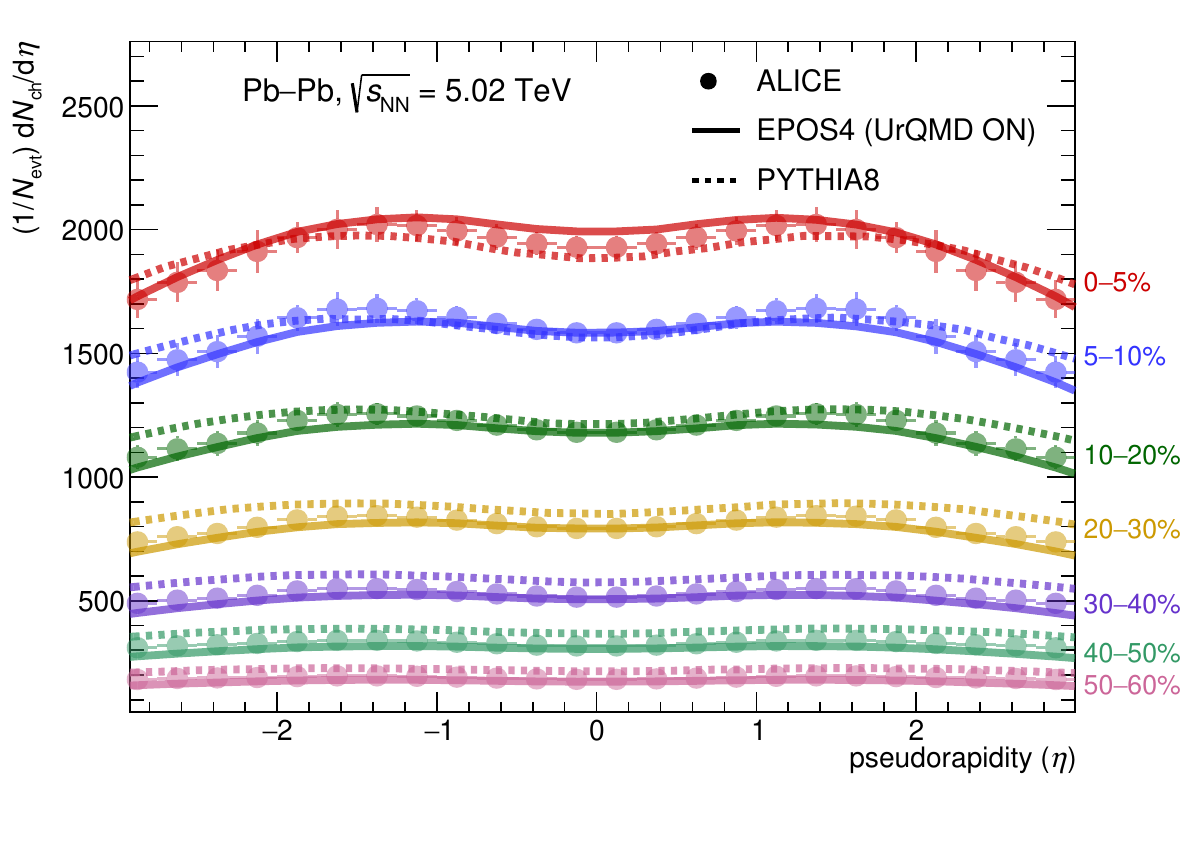}
\caption{Charged-particle pseudorapidity density distributions normalized with number of events from Pb--Pb collisions at \energy{5.02} obtained using EPOS4 (solid lines) and PYTHIA8+Angantyr (dotted lines) across various centralities are shown. The ALICE data (filled markers)~\cite{ALICE:2016fbt} is also shown.}
\label{figure4}
\end{figure}
\begin{figure}[htp]
\centering
\includegraphics[scale=0.45]{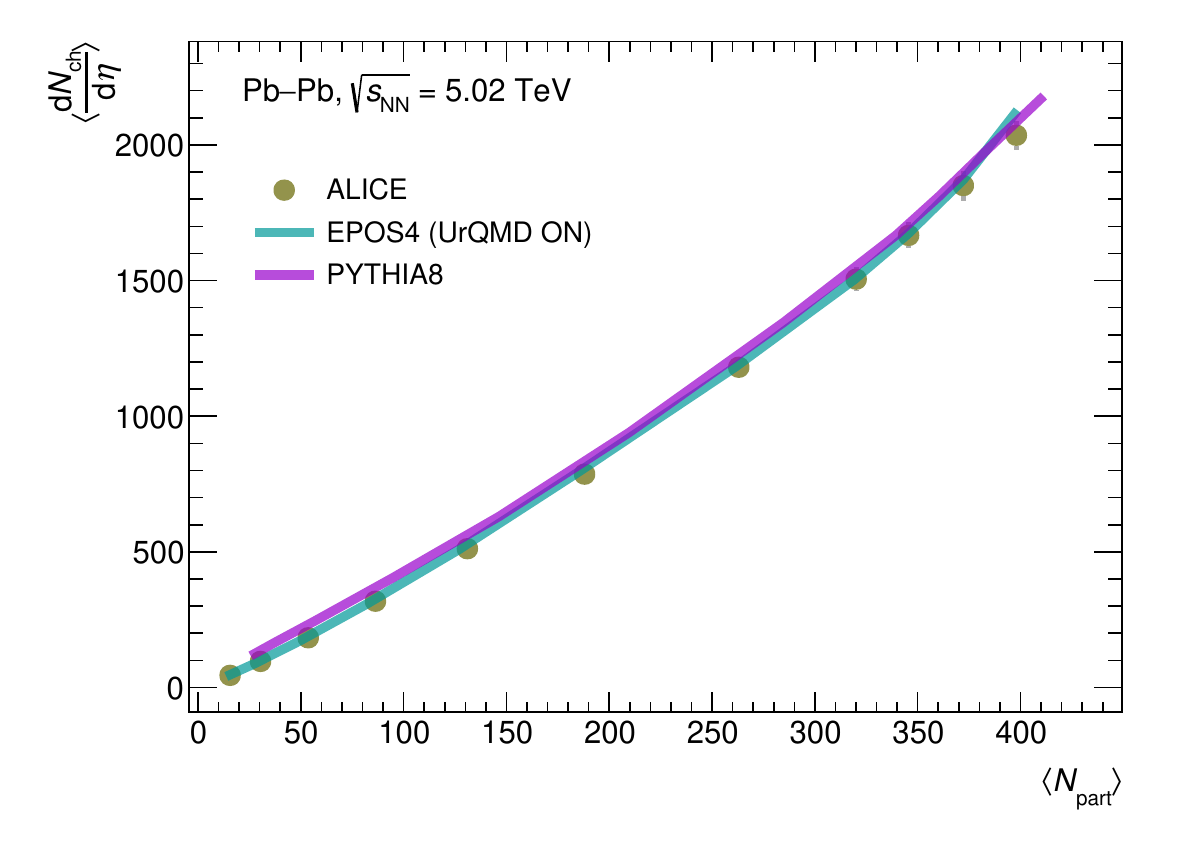}
\caption{Mean charged particle density  $\langle d\mtwo{N}{ch}/d\eta \rangle$ as a function of  participant nucleons (\two{N}{part}) for EPOS4 and PYTHIA8+Angantyr, compared with ALICE measurements\cite{ALICE:2015juo} for Pb--Pb collisions at \energy{5.02}.}
\label{figure5}
\end{figure}
\subsection{Event samples}
Events for the analysis are generated for the Pb--Pb collisions at \energy{5.02} using the above mentioned Monte Carlo event generators. The characteristics of collision events are strongly influenced by the collision centrality, which quantifies the initial geometric overlap of the colliding nuclei and is closely related to the event activity in heavy-ion collisions. Geometry proxies such as impact parameter $(b)$, participant nucleons (${N}_{part}$) and binary collisions (${N}_{coll}$) are not directly measurable in experiment and are typically inferred by correlating event activity (e.g. charged‑particle yield or summed transverse energy) with a Glauber‑based geometry~\cite{Miller:2007ri}. In simulations, centrality can be defined either directly from the generated $b$ or via midrapidity multiplicity/summed transverse energy within a chosen $\eta$ window. For PYTHIA8, centrality has been estimated using the distribution of the summed transverse energy ($\sum E_{\mathrm{T}}$) in the final state. Figure~\ref{figure3} presents the $\sum E_{\mathrm{T}}$ distribution, where the percentile boundaries defining the same centrality classes as used for EPOS4 (Fig.~\ref{figure2}) are indicated. This allows a consistent comparison between EPOS4 and PYTHIA8 by mapping the event activity in both models to a common centrality framework. Centrality for the EPOS4 simulated events is determined using $b$ where Figure~\ref{figure2} shows the number of events as a function of $b$. The distribution is divided into several percentiles based centrality classes corresponding to the most central to peripheral collisions. These intervals provide a systematic way to relate the geometric overlap of the colliding nuclei to event activity within the model.\\
Figure \ref{figure4} shows the normalized charged-particle pseudorapidity density distributions~\cite{Franceschini:2022vck}, $(1/N_{\mathrm{evt}})\, dN_{\mathrm{ch}}/d\eta$, for different centrality classes, comparing EPOS4 (solid lines) and PYTHIA8 (dotted lines) with ALICE experimental data (filled markers)~\cite{ALICE:2016fbt} at $\sqrt{s_{\mathrm{NN}}} = 5.02~\mathrm{TeV}$. The results are presented for centrality intervals $0$--$5\%$, $5$--$10\%$, $10$--$20\%$, $20$--$30\%$, $30$--$40\%$, $40$--$50\%$, and $50$--$60\%$. Simulated events from the two models are observed to have generated tracks with $\eta$ density very close to ALICE data in the mid-rapidity region. Figure~\ref{figure5} shows the mean charged-particle pseudorapidity density at midrapidity ($\abs{\eta} < 0.8$) as a function of participant nucleons, for the events obtained from the two models and the ALICE data points. Event samples generated using PYTHIA8 and EPOS4 (UrQMD ON and UrQMD OFF) are observed to explain the mean charged particle pseudorapidity density from ALICE for Pb--Pb collisions at \energy{5.02}. These two figures give a baseline characterization of the simulated events with regard to ALICE data and validate the consistency of event generation and the centrality determination of the events in the subsequent analysis.
\begin{figure} [h!]
\includegraphics[scale= 0.45]{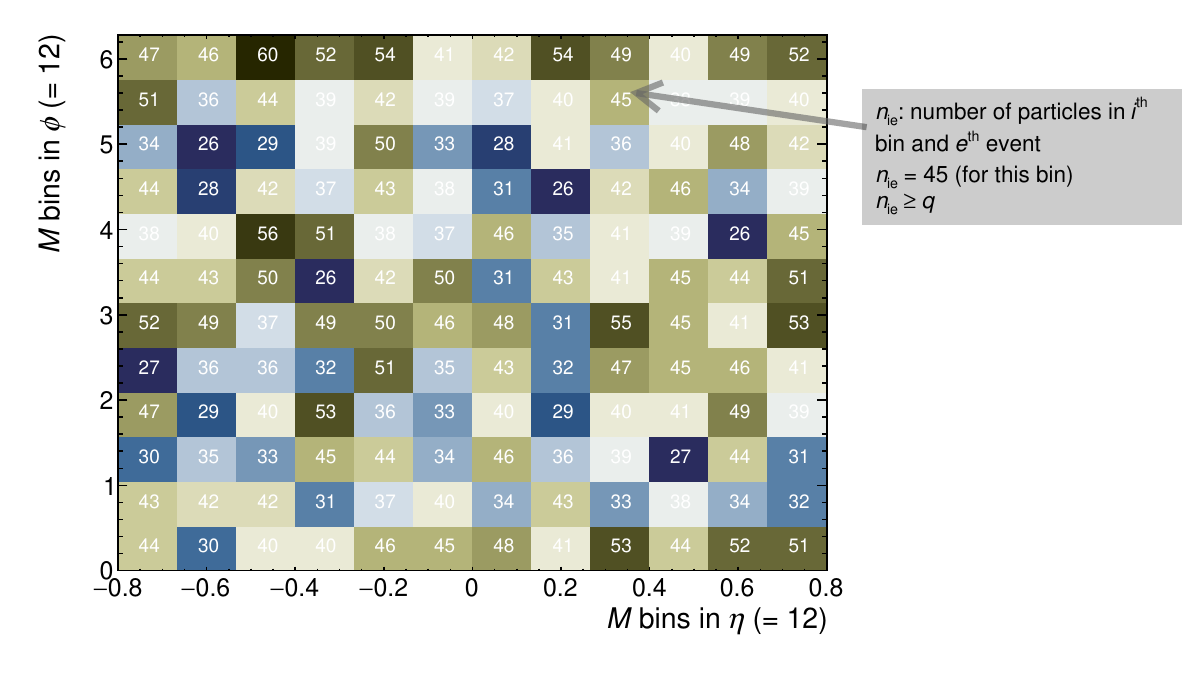}
\caption{\label{fig:grid} Schematic illustration of the two-dimensional phase-space partitioned into equal number of bins ($M$) along $\eta$ and the $\varphi$ direction. The numbers inside the cells represent the particle count, \protect$n_{ie}\protect$ termed as the bin multiplicity.}
\end{figure} 
%
%%%%%%%%%%%%%%%%%%%%%%%%%%%%%%%%%%%%%%%%%%%%%%%%%%%%%%%%%%%%%%%%%%%%%%%
\section{Analysis and Methodology} \label{sec:Methodology}   
Event-by-event fluctuations in particle densities carry valuable information about the underlying dynamics and may reveal signatures of critical phenomena associated with the transition between hadronic matter and QGP\cite{Koch:2001zn}. Local charged-particle density fluctuations, quantified through normalized factorial moments (NFMs) and intermittency observables, provide a sensitive probe of critical dynamics and the multifractal structure of particle production in relativistic heavy-ion collisions~\cite{Bialas:1985jb,Hwa:2016khr}. It provides a framework to study local density fluctuations by examining the scaling behaviour of NFMs with decreasing phase-space intervals. A non-trivial power-law dependence indicates self-similar structures in particle production and can serve as evidence of fractal behaviour. Such fluctuations are scale invariant and are particularly interesting because they are expected near critical point, where long-range correlations and density fluctuations become enhanced~\cite{Antoniou:2017vti,Hwa:1992uq}. 
\par
Intermittency analysis first introduced in 1986~\cite{Bialas:1985jb,Bialas:1988wc} in the field of heavy-ion collisions has been performed for the study of various collision systems during the late 1980s and throughout the 1990s, to reveal features of particle production. Investigations across a wide range of collision systems, including hadron–hadron, hadron–nucleus, nucleus–nucleus and cosmic-ray interactions, have been performed%Experimental studies performed by collaborations such as NA22, NA27, UA1 and nuclear-emulsion experiments reported scaling behaviour of normalized factorial moments in rapidity and multidimensional phase spaces
~\cite{DeWolf:1995nyp}. More recently, with the availability of high-statistics data from collider experiments at RHIC and the LHC, intermittency studies have regained attention in the context of searching for critical behaviour and understanding the fractal nature of particle production at high energy densities. Substantially larger multiplicities available at modern collider energies have also enabled multidimensional analyses with improved statistical precision, allowing a more detailed study of scale-invariant structures in particle production dynamics. Various experimental~\cite{ReynaOrtiz:2024hul,Marcinek:2017uzp,STAR:2023jpm,Sharma:2023ndr,Malik:2024ltm} and model studies~\cite{Sharma:2018vtf,Gupta:2019zox,Haider:2026rtw,Sarma:2019teo,Singh:2024gai} have been performed in this direction. The methodology~\cite{Hwa:2011bu} adopted in the present work for investigating the scaling behaviour of particles produced in heavy-ion collisions is detailed below.
\begin{figure}[h!]
    \includegraphics[scale=0.45]{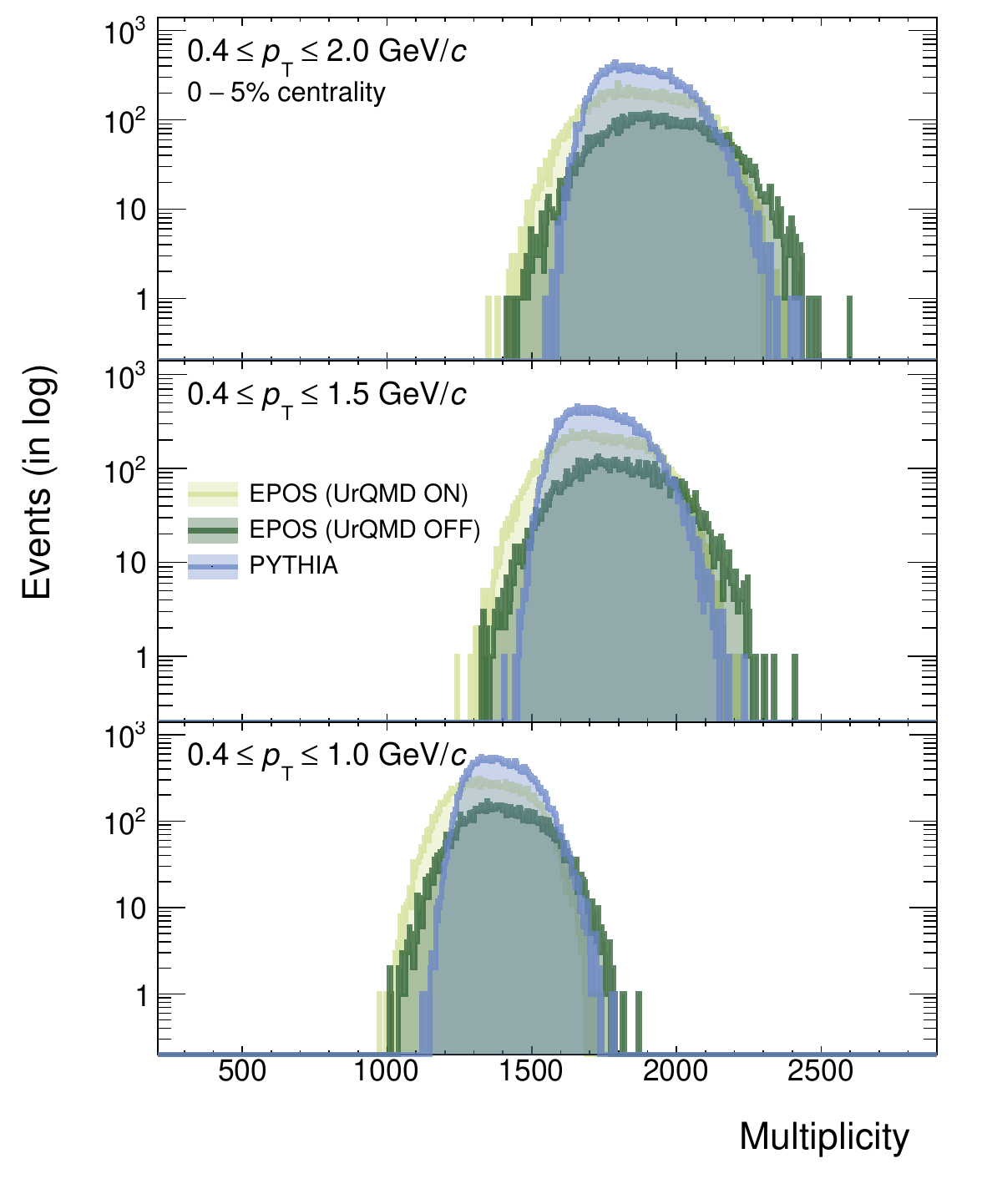}
    \caption{Charged‑particle multiplicity distributions within $|\eta| \leq 0.8$ for 0--5\% central Pb--Pb collisions events at \energy{5.02} for the \two{p}{T} ranges: \ptrange{0.4}{2.0}~GeV/c(top), \ptrange{0.4}{1.5}~GeV/c (middle), \ptrange{0.4}{1.0}~GeV/c (bottom) shown on a logarithmic event count scale.}
    \label{figure6}
\end{figure}

\begin{figure*}[htb]
    \includegraphics[width=0.3\textwidth]{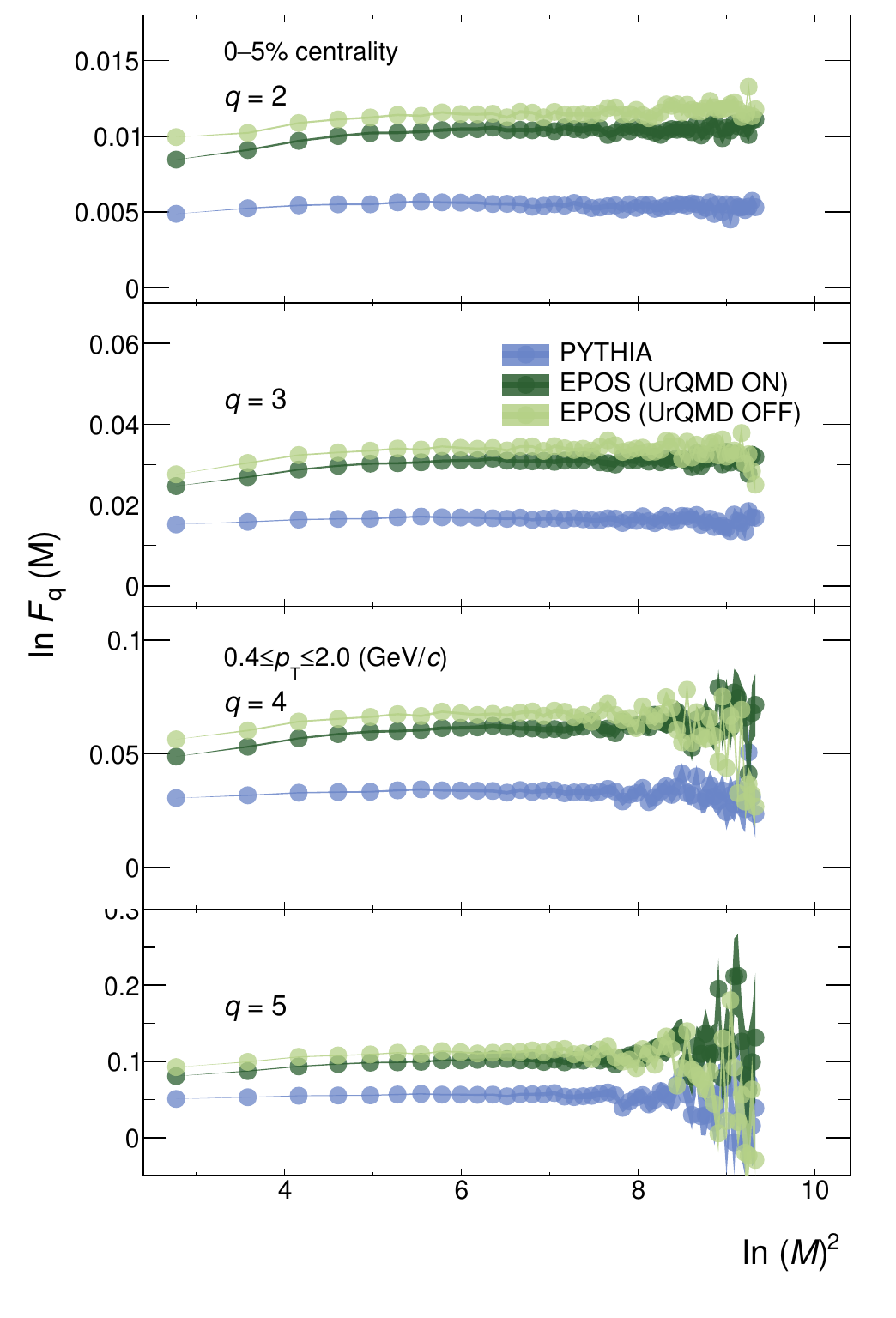} 
    %\hspace{-4mm}
    \includegraphics[width=0.3\textwidth]{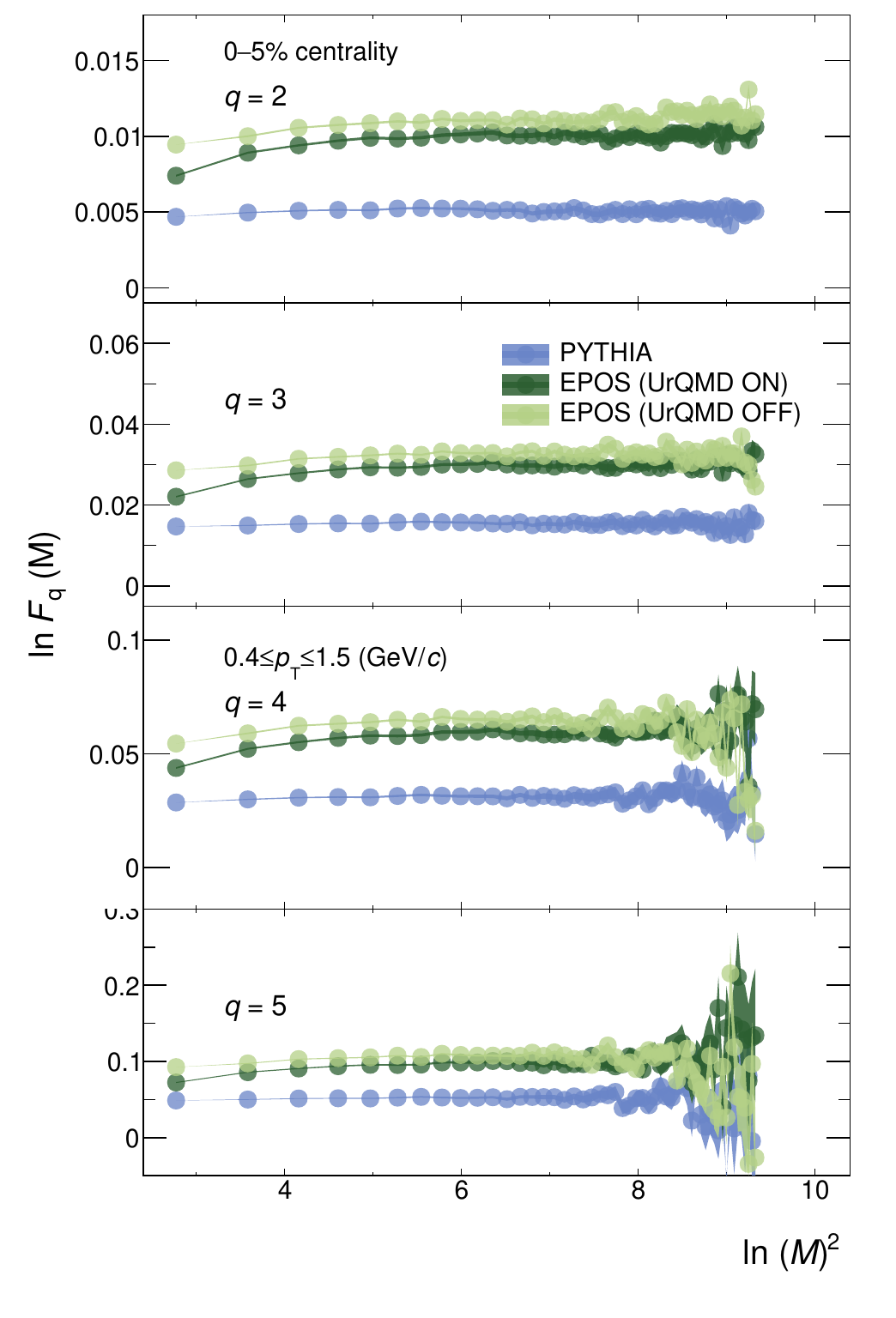}
    %\hspace{-4mm}
    \includegraphics[width=0.3\textwidth]{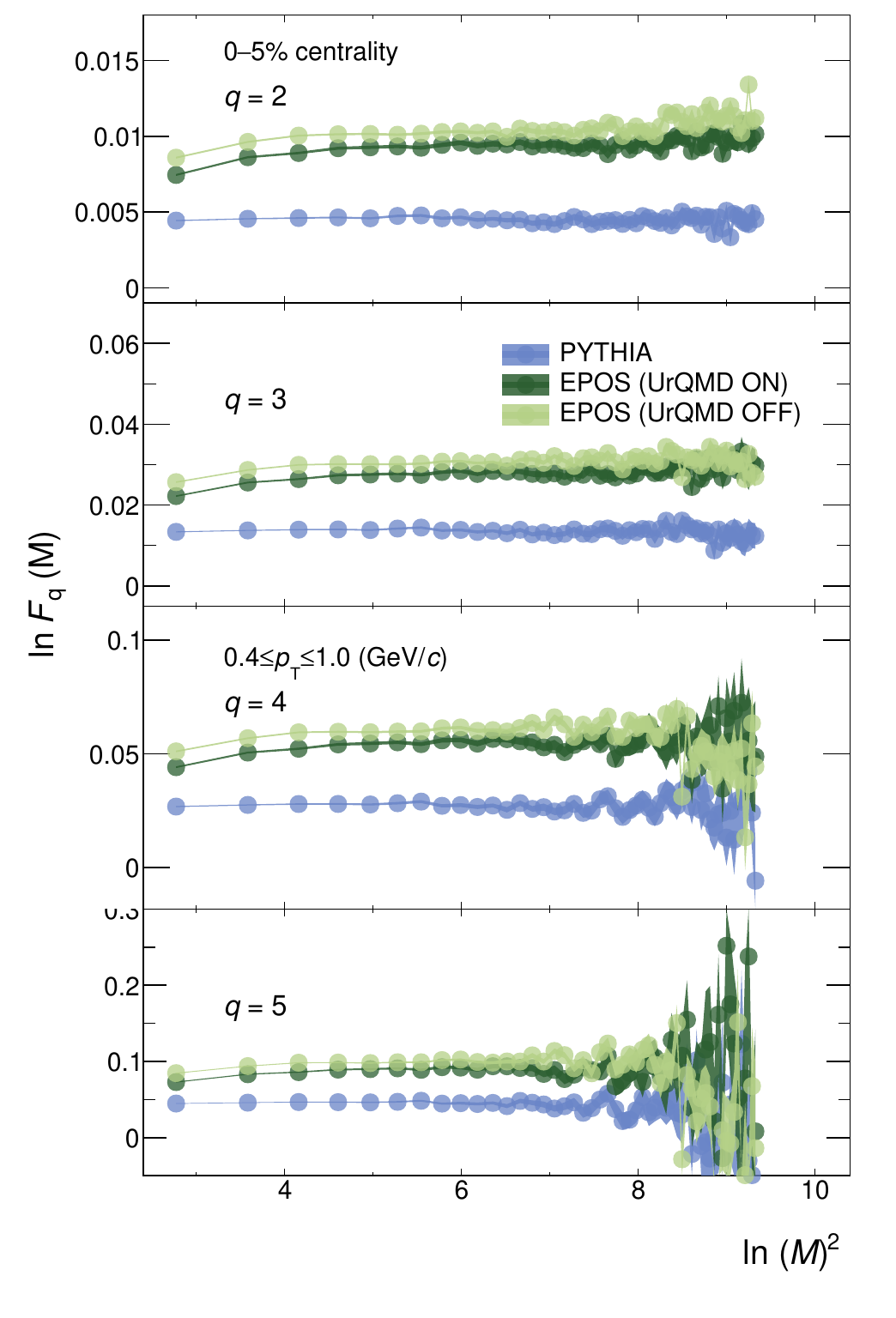}
    \caption{\label{fig6} M--scaling: ln \fq{q} as a function of  ln~\sqm~  for 0--5\% central events showing results and comparison of PYTHIA8, EPOS4  for UrQMD ON and  UrQMD OFF modes across \two{p}{T} ranges: \ptrange{0.4}{2.0} (left), \ptrange{0.4}{1.5} (middle), \ptrange{0.4}{1.0} (right) for $q$ = 2, 3, 4 and 5.}.
    \label{figure7}
\end{figure*}
\begin{figure}[h!]
    \includegraphics[width=0.45\textwidth]{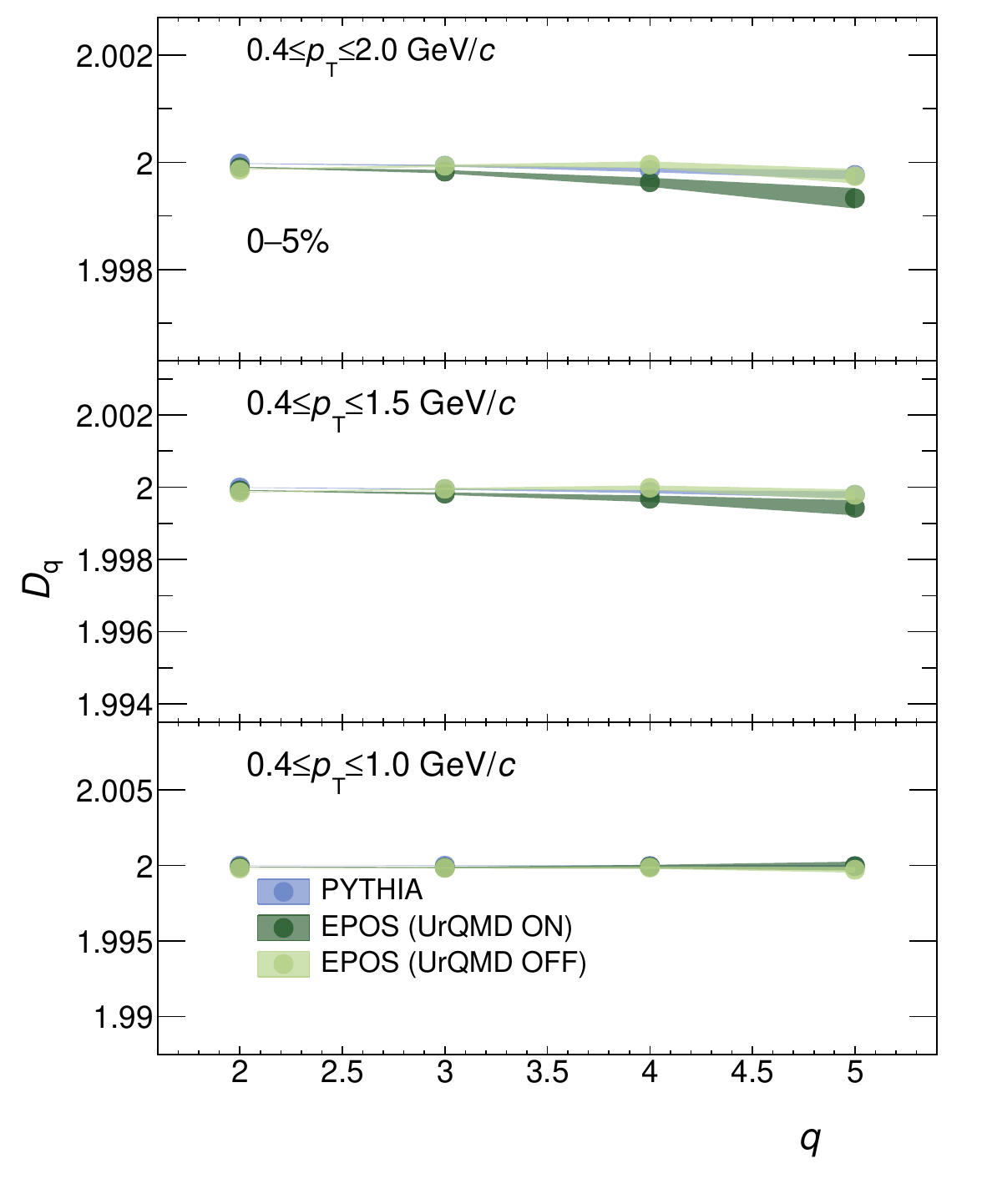} 
    \caption{\label{figure10} Generalized (Rényi) dimensions, \two{D}{q} as a function of $q$ for \two{p}{T} ranges: \ptrange{0.4}{2.0}~GeV/c (top), \ptrange{0.4}{1.5}~GeV/c (middle), \ptrange{0.4}{1.0}~GeV/c (bottom). Observations are for the charged particles in the 0--5\% central events generated using PYTHIA8 and EPOS4 with UrQMD ON and OFF. }.
\end{figure}
%%%%%%%%%%%%%%
The scaling behaviour of the particles produced in two dimensional angular phase space $(\eta, \varphi)$ of an event are studied. For an event particles are distributed over $(\eta, \varphi)$ grid with $M$ bins along the $\eta$ axis and $M$ bins along the $\varphi$ axis, as shown in Figure~\ref{fig:grid}.  Each bin in the grid, with ${M^2}$ cells, represents a specific region of the phase space, and $n_{\rm{ie}}$ indicates the number of particles in the $i^{\rm{th}}$ bin of the $e^{\rm{th}}$ event. The event factorial moment of order $q$ is calculated as
\begin{equation}
 f_{q}^{e}(n_{\rm{ie}}) = \displaystyle\prod_{j=o}^{q-1}(n_{\rm{ie}} - j).
\label{equation2}
\end{equation}
For $N$ events, the normalised factorial moments~\cite{Bialas:1988wc, Bialas:1985jb}, which serve as a measure of departure from a Poissonian (random) distribution are  determined as:
\begin{equation}
  F_{\rm{q}}(M)= \displaystyle \frac{\frac{1}{N} \displaystyle\sum_{e=1}^{N}\frac{1}{M^2}\sum_{i=1}^{M}f_{q}^{e}(n_{\rm{ie}})}{\left (\frac{1}{N} \displaystyle \sum_{e=1}^{N}\frac{1}{M^2}\sum_{i=1}^{M}f_{1}^{e}(n_{\rm{ie}}) \right )^q}. 
\label{equation1}
\end{equation}
\par
For a system that exhibits dynamical fluctuations associated with critical behaviour, the normalized factorial moments  $F_q(M)$ are expected to increase systematically as the number of phase-space partitions (M) grows, giving linear dependence. This behaviour also reflects self-similar fluctuations in the system and is commonly referred to as M-scaling, with the corresponding growth characterized by an index called the  intermittency index $\phi_{\rm{q}}$ such that: 
 \begin{equation} 
F_{\rm{q}}(M) \propto M^{\phi_{q}}.
\label{def}
\end{equation} 
The parameter $q$ is a positive integer greater than 1. A strict linear dependence of ${F_{q}}$ on M is indicative of the presence of critical fluctuations.  However, on the other extreme, no dependence of $F_{q}$ on M  is understood to be due to pure Poisson nature of particle distribution. 
\par
An important aspect of multiparticle production is the possible emergence of self-similar structures and scale-invariant behaviour. Such properties naturally connect with the mathematical framework of fractals. Fractal concepts have appeared in various branches of physics where systems display patterns that repeat across different scales. In high-energy physics, a cascade process during parton evolution and hadronization may generate structures exhibiting similar characteristics. Fractals are structures that exhibit repeating patterns over different scales and have  fractional dimension different from their topological dimension, reflecting their inherent complexity and irregular geometry. Fractal dimensions and intermittency index are mathematically related as~\cite{Sarkisian:1993gz,DeWolf:1995nyp}: 
\begin{equation} 
{ D_{q} = D_{T}~(1- \frac{\phi_{q}}{q-1})}.
\label{eq:defdq}
\end{equation}
A constant  $\rm{D}_{q}$ with $q$ suggests monofractal behaviour, while a dependence of $\rm{D}_{q}$ on $\rm{q}$ indicates multifractality. Multifractal structures imply that different regions of phase space exhibit different scaling properties, which, if present,  reflects complex dynamical evolution during hadronization. Whereas, fluctuations exhibit different scaling behaviours for different scales or regions~\cite{Kamal:2015rxa} in case of multifractal system. In a statistically homogeneous fractal distribution~\cite{Sarkisian:1993pi}, the properties of particle distributions remain invariant under scaling transformations, ensuring that the probability of detecting particles within a given volume is consistent across different regions of phase space. The power-law behaviour associated with a homogeneous fractal dimension then exhibits self-similarity across multiple scales. This fundamental property is subsequently reflected in the scale-invariant behaviour of the NFM. 

%%%%%%%%%%%%
 \begin{figure*}[htb]
    \includegraphics[width=0.95\textwidth]{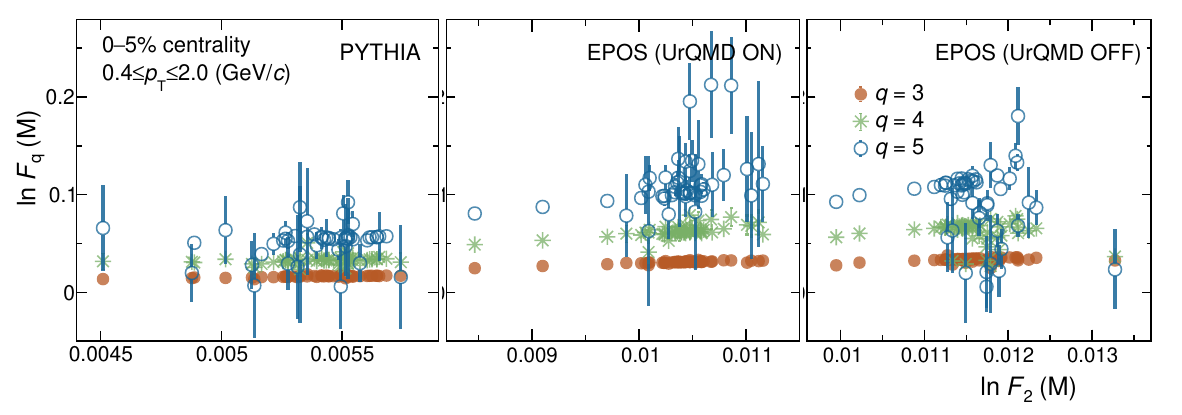}\\ 
    %\hspace{-4mm}
    \includegraphics[width=0.95\textwidth]{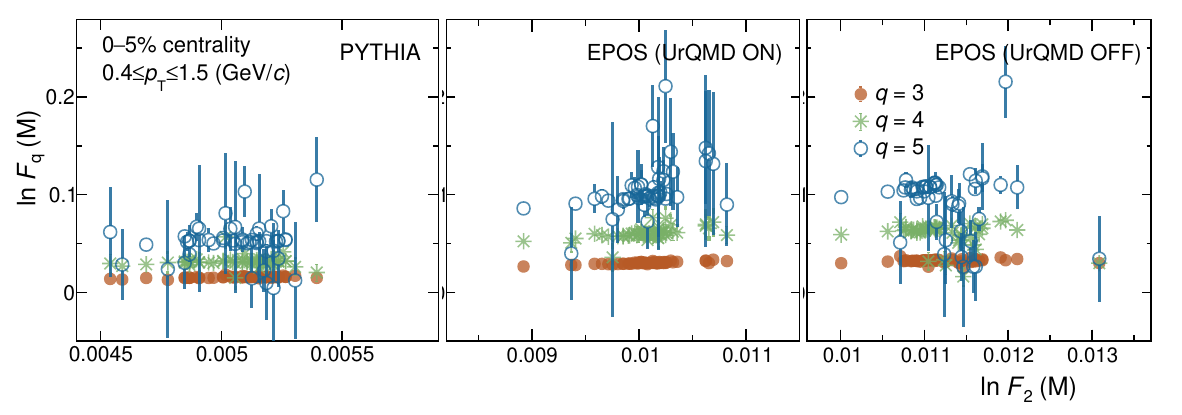}\\
    %\hspace{-4mm}
    \includegraphics[width=0.95\textwidth]{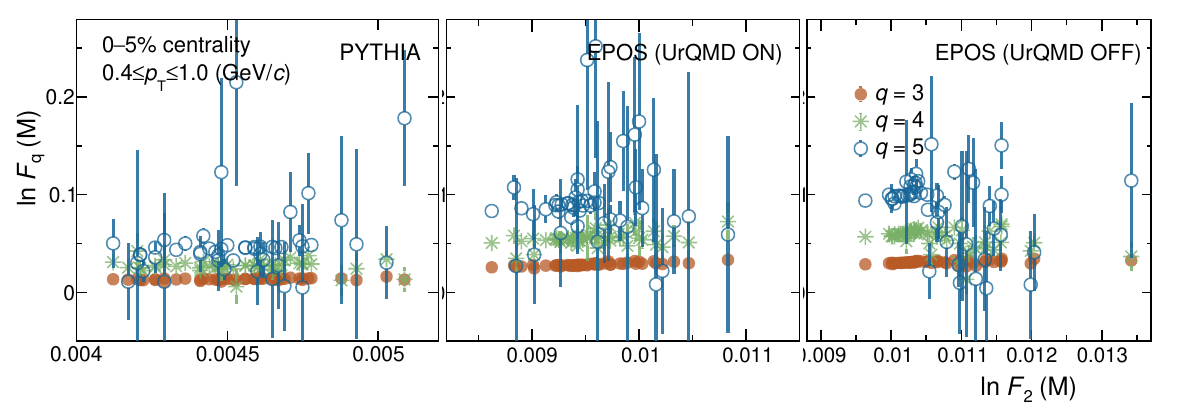}\\
    \caption{\label{fig6} ln (\fq{q}) as a function of ln (\fq{2}) (F--scaling) for PYTHIA8, EPOS4 UrQMD ON and OFF modes for \two{p}{T} ranges: \ptrange{0.4}{2.0}~GeV/c (top), \ptrange{0.4}{1.5}~GeV/c (middle), \ptrange{0.4}{1.0}~GeV/c (bottom) for the most central 0--5\% events. }.
    \label{figure8}
\end{figure*}

\begin{figure*}[htb]
    \includegraphics[width=0.99\textwidth]{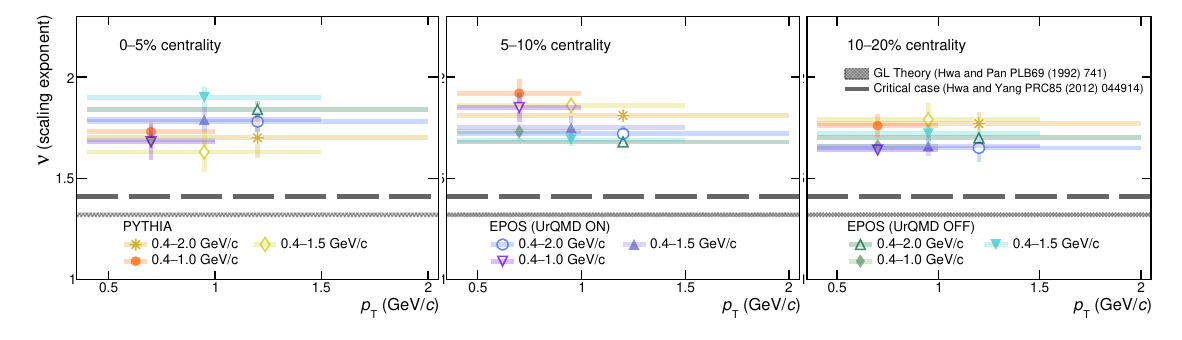}
    \caption{\label{figure9} Scaling exponent ($\nu$) as a function of \two{p}{T} for \ptrange{0.4}{2.0}~GeV/c, \ptrange{0.4}{1.5}~GeV/c, \ptrange{0.4}{1.0}~GeV/c and  0--5\% (left), 5--10\% (middle), 10--20\% (right) central events. }.
\end{figure*}
%%%%%%%%%%%%

\par 
Further, in the transition of a system from one phase to another, there can be different pathways. Order scaling is observed to be present in the systems going through second order phase transition~\cite{Hwa:1992uq} such that:
\begin{equation} 
F_{\rm{q}}(M) \propto(F_{2}(M))^{\beta_{q}}.
\label{def}
\end{equation} 

This power-law behaviour is referred to as F-scaling~\cite{Hwa:1992uq}. The exponent $\beta_{q}$ characterizes the scaling behaviour of the higher-order factorial moments with respect to the second-order moment. The power-law relationship between $\beta_{q}$ and $q$ is given by the relation: 
\begin{equation} 
{\beta_{q} \propto (q-1)^\nu}
\label{eq:defnu}
\end{equation}
that characterizes scaling behaviour across different moment orders and remains independent of the dimensionality of the phase space bins where $\nu$, the scaling exponent is a dimensionless exponent.

%%%%%%%%%%%%%%%%%%%%%%
\section{Observations and Results}
Event  samples of Pb--Pb collisions at $\sqrt{s_{\mathrm{NN}}}=5.02~\mathrm{TeV}$ are generated using PYTHIA8 (200K) and  EPOS4 (300K) Monte Carlo event generators. Charged particles ($\pi^{\pm}, ~K^{\pm},~p,~\overline{p}$) in the central events within the kinematic acceptance $|\eta| \leq 0.8$, full azimuth ($0 \leq \varphi \leq 2\pi$) in the soft \two{p}{T} intervals \ptrange{0.4}{2.0}, \ptrange{0.4}{1.5}, \ptrange{0.4}{1.0}~GeV/$c$ are studied to investigate the  scaling and fractal behaviour in these event generators. The multiplicity distributions of the generated charged particles within the acceptance bounds, from PYTHIA8, EPOS4 (UrQMD ON) and EPOS4 (UrQMD OFF) are given in Fig.~\ref{figure6}. For all the $p_{\mathrm{T}}$ bins,  distributions are observed to be well described by the Gaussian shapes with no secondary peaks.
\par
As discussed in section~\ref{sec:Methodology}, the charged particles in an event are mapped onto the two-dimensional angular $(\eta, \varphi)$ phase-space grid. The phase space is partitioned into M\,x\,M bins, where the partition number is varied from 4 to 82 in steps of 2. The factorial moments, defined for each bin (Eq.~(\ref{equation2})) are evaluated for q = 2 - 5, for each partition. The above procedure is repeated for all events to determine event factorial moments  that are subsequently averaged over the entire event sample to obtain the normalized factorial moments (\fq{q}), as defined in Eq.~(\ref{equation1}). 
\par
The behaviour of $\ln F_{\rm{q}}(M)$  as function of $\ln\,(M^{\rm{2}})$ for order q = 2, 3, 4, and 5 in the 0-5\% most central events within the  three ${p_{\rm{T}}}$ intervals (\ptrange{0.4}{2.0} GeV/c (left), \ptrange{0.4}{1.5} GeV/c (middle), \ptrange{0.4}{1.0} GeV/c (right)) are shown in Figure~\ref{figure7}. The same qualitative scaling behaviour and systematic hierarchy of the normalized factorial moments are observed across all three $p_{\mathrm{T}}$ intervals and the other two centrality classes studied.  The results are compared for the three Monte Carlo event generators and for all these it is observed that  $\ln F_{\rm{q}}(M)>0$  and \fq{q+1}$\geq$\fq{q}, reflecting the increasing sensitivity of higher-order moments to fluctuations. Further $F_{\rm{q}}(M) \sim 1$  initially increases with increasing phase-space partitioning, followed by a gradual saturation observed for all orders of q. The saturation is most prominent in PYTHIA8 in comparison to EPOS4. Similar observations are made across the three $p_{\mathrm{T}}$ bins. This independence of $F_{\rm{q}}(M)$ on the increasing number of bins implies weak bin-to-bin fluctuations in the local particle number density, scale dependent charged particle generation and thus absence of intermittency in these models. In addition this saturation can arise because of finite particle multiplicity, finite correlation length or because of dynamics of the model. PYTHIA8 consistently yields lower values of $\ln F_{\rm{q}}(M)$ than that in EPOS4 for all orders, indicating comparatively weaker fluctuations in it. The inclusion of the hadronic cascade (UrQMD ON) generally produces slightly larger values than the corresponding UrQMD OFF calculation, particularly for higher orders (q=4 and q=5), suggesting that hadronic rescattering enhances dynamical fluctuations. 
\par
The vertical bars on markers in the figure denote statistical uncertainties obtained using sub‑sampling method. For q = 4 and especially q = 5, the statistical uncertainties increase noticeably at the highest partition numbers ($\ln (M^2)~ \ge~8.5 $). This behaviour is expected in case the average particle occupancy per bin becomes very small at large M, causing higher-order factorial moments to become increasingly sensitive to statistical fluctuations and limited event statistics.     
 \par
Line fits were performed in the high-M region of the $\ln F_{\rm{q}}(M)$ versus $\ln (M^{2})$ graphs for each order $q$. The slope values, the intermittency indices (\two{\phi}{q}), obtained  from these fits are used to determine the generalized (Rényi) dimensions \two{D}{q}, with the phase‑space dimensionality $D_{\rm{T}}=2$. Fig.~\ref{figure10} shows variation of \two{D}{q} as a function of the order   $q$ for the three ${p_{\rm{T}}}$ intervals in $0-5\%$ central collisions. The results obtained from PYTHIA8, EPOS4 (UrQMD ON) and EPOS4 (UrQMD OFF) exhibit similar overall trend across all $p_{\rm{T}}$ intervals. Within uncertainties (shown as coloured bands), \two{D}{q} exhibits negligible $q$-dependence. This behaviour indicates that the charged-particle production simulated by all three event generators is predominantly monofractal in nature for central heavy-ion collisions, with no significant evidence of multifractality.
\par
The higher-order normalized factorial moments, $\ln F_{\rm{q}}(M)$ for q = 3, 4, and 5, were studied as a function of the second-order normalized factorial moment, ($F_{\rm{2}}(M)$), to investigate the F-scaling behaviour described by Eq.~\ref{def}. Figure~\ref{figure8}  presents the dependence of $\ln F_{\rm{q}}(M)$ on $F_{\rm{2}}(M)$ for charged particles generated using PYTHIA8, EPOS4 (UrQMD ON) and EPOS4 (UrQMD OFF) in the 0–5\% centrality class for the investigated transverse momentum intervals. The observed trends remain consistent across all three $p_{\rm{T}}$ intervals and the other two centrality classes considered. For all three models,  $F_{\rm{q}}(M)$ with q = 3, 4 and 5 exhibits only a very gradual increase with increasing $F_{\rm{2}}(M)$. This increase is least pronounced for PYTHIA8 and both EPOS4 calculations show a comparatively stronger dependence, particularly for the higher-order moments. Overall, the three models exhibit an approximately linear F-scaling behaviour over the studied range.
\par
The F-scaling exponent, $\beta_{\rm{q}}$, for q = 3, 4, and 5 is determined from linear fits to the F-scaling plots. Subsequently, the scaling exponent ($\nu$) is extracted using Eq.~\ref{eq:defnu}. Figure~\ref{figure9} presents the variation of $\nu$ with transverse momentum for the three centrality classes, 0–5\%, 5–10\% and 10–20\%. The values of scaling exponent $\nu$  to examine the dependence of  scaling behaviour on collision centrality and transverse momentum are presented with observations from PYTHIA8, EPOS4 (UrQMD ON), and EPOS4 (UrQMD OFF).   Horizontal bars on the markers represent the transverse momentum ($p_{\rm{T}}$) intervals corresponding to each data point. For comparison, the predicted values of the $\nu$ from the Ginzburg–Landau (GL) theory~ \cite{Hwa:1992uq} and the SCR model~\cite{Hwa:2011bu}, expected for a system undergoing a second-order phase transition, are also shown. The values of ($\nu$) obtained from PYTHIA8, EPOS4 (UrQMD ON) and EPOS4 (UrQMD OFF) lie in the range ($1.6 \lesssim \nu \lesssim 1.9$) for all investigated centrality classes and $p_{\rm{T}}$ intervals. These values are consistently higher than the GL theory and SCR model predictions This implies  that the particle production mechanisms implemented in these event generators do not exhibit the scaling behaviour expected for a system at or near the critical point associated with formalism for second-order phase transition under GL theory or as implemented in the SCR model.
\section{Summary} \label{sec:Summary}
The present study investigates the scaling behaviour of charged-particle multiplicity fluctuations in Pb–-Pb collisions at $\sqrt{s_{\mathrm{NN}}}=5.02$ TeV using simulated event samples generated with EPOS4 (UrQMD ON and UrQMD OFF) and PYTHIA8. The local multiplicity fluctuations are probed using normalized factorial moments to examine sensitivity of intermittency observables to different particle-production mechanisms. Only a weak increase in the factorial moments  followed by saturation at finer phase-space partitions is observed, indicating the absence of a clear power-law scaling over the investigated range and hence absence of M-scaling. The generalized (Rényi) dimensions exhibit little dependence on the order of the moment, suggesting a predominantly monofractal nature of the simulated particle generation. Furthermore, the extracted scaling exponent ($\nu$) lies in the range $\sim$ 1.6 to 1.9, consistently exceeding the values predicted by the models for a second-order phase transition. Although PYTHIA8+Angantyr, EPOS4, and EPOS4 with UrQMD exhibit qualitatively similar dependence of the normalized factorial moments on phase-space resolution, their absolute magnitudes differ systematically. The vertical displacement of the $\ln F_{\rm{q}}(M)$   distributions, with EPOS4+UrQMD > EPOS4 > PYTHIA8+Angantyr, indicates a progressive enhancement of local dynamical multiparticle correlations due to collective medium evolution and hadronic rescattering. However, the preservation of the overall scaling behaviour demonstrates that these additional physical processes modify only the strength of the correlations and not their scale dependence. Consequently, neither hydrodynamic evolution nor hadronic rescattering as implemented in these models generate the self-similar, scale-invariant fluctuations required for intermittency under the present conditions.

\section*{Acknowledgements} \label{sec:acknowledgements}
%\lipsum[8]
RG gratefully acknowledge financial support for this work from the Dean Research Studies, University of Jammu and from the Department of Science and Technology (DST), Government of India, via the project “Indian participation in ALICE experiment at CERN” sanctioned under order No.~3015/1/2021/Gen/R\&D-I/13283. One of the authors (FUH) acknowledges the University of Jammu and the Council of Scientific and Industrial Research (CSIR), Government of India, for financial support through a research fellowship. PC acknowledges the DST, Govt. of India for financial support through DST-Inspire fellowship. The authors also thank the CERN lxplus computing grid for the computational resources used partially for this analysis.
%  \lipsum[6-7]

%\begin{thebibliography}{4} 
\bibliographystyle{unsrt}
\bibliography{main}
%\end{thebibliography}

%\input{sections/appendix1.tex}

\end{document}